# Photonically-confined solar cells: prospects for exceeding the Shockley-Queisser limit

Qian Zhou[1], Arfa Karani[2], Yaxiao Lian[1], Baodan Zhao[1,2], Richard H. Friend[2], Dawei Di[1,2]*

1. State Key Laboratory of Modern Optical Instrumentation, College of Optical Science and Engineering; International Research Center for Advanced Photonics, Zhejiang University, Hangzhou, 310027, China.
2. Cavendish Laboratory, University of Cambridge, JJ Thomson Avenue, Cambridge, CB3 0HE, United Kingdom.

* Corresponding author, Email: daweidi@zju.edu.cn

## Abstract

The Shockley-Queisser (SQ) limit[1], introduced by W. Shockley and H. J. Queisser in 1961, is the most well-established fundamental efficiency limit for single-junction photovoltaic solar cells. For widely-studied semiconductors such as Si, GaAs and lead-halide perovskite, the SQ limits under standard solar illumination (1-sun) are 32.7%, 32.5% and 31% for bandgaps of 1.12 eV, 1.43 eV and 1.55 eV, respectively[2]. Here, we propose that the fundamental efficiency limits for single-junction solar cells may be surpassed via photon confinement, substantially raising the theoretical limits to 49%, 45.2% and 42.1% for Si, GaAs and methylammonium lead iodide ($MAPbI_3$) perovskite cells under 1-sun. Such enhancement is possible through the containment of luminescent photons within the solar cell, allowing the suppression of both non-radiative and radiative recombination losses, which were considered inevitable in the classical SQ model. Importantly, restricting photon emission from the solar cells raises the open-circuit voltage ($V_{OC}$) to values approaching the semiconductor bandgaps, surpassing the theoretical $V_{OC}$ values predicted by the SQ model. The fill factors of the cells are expected to increase substantially, resulting in current-voltage characteristics with very-high squareness for ideal diode operation. Our work introduces a new framework for improving solar cell performance beyond the conventional limits.

**Main Text**

Energy generation by photovoltaic solar cells has become an essential technology for the sustainable development of human society. While silicon solar cells[3,4] have been the dominant driving force for commercial applications, photovoltaic cells employing III-V semiconductors[5,6] offer superior energy-conversion efficiencies. Emerging photovoltaic technologies[7-9], including thin-film chalcogenide[10-13], organic[14-17], dye-sensitized[18,19], and more recently, perovskite[20-29] solar cells, promise low-cost solar electricity generation. However, the performance of single-junction solar cells made using any materials is well-established to be fundamentally constrained by the Shockley-Queisser (SQ) model[1], a theoretical hallmark for photovoltaics originally published in 1961. While tandem solar cell structures are the only practical solution to date amongst a number of next-generation photovoltaic concepts for circumventing this limitation[8,30-32], there are benefits for achieving ultra-high efficiencies using single-junction solar cells, which have reduced processing complexity and less-stringent requirement for spectral matching.

One of the most important assumptions behind the SQ model is that non-radiative recombination losses must be completely eliminated to reach the ultimate efficiency limit[1], while radiative recombination losses cannot be avoided, leading to an open-circuit voltage (or quasi-Fermi level splitting) limit considerably lower than the semiconductor bandgap. In this work, we consider the possibility that radiative recombination pathways can be restricted via photon confinement, raising the quasi-fermi-level splitting to an ideal maximum approaching the semiconductor bandgap. In contrast to the established concepts of photon management and recycling[26,27] in which the ultimate emission efficiency approaches the SQ limit, our model presents an opportunity of greatly exceeding the limit with a single-junction cell without employing concentrations above 1 sun or angular emission restrictions.

To conceptualize a photonically-confined solar cell (PCSC), we consider an ideal solar cell model that generally follows the original SQ formulation[1], with the exception that each photon created within the solar cell as a result of luminescence has a finite probability of escaping the cell. The probability of escape ($f_{esc}$) is normalized to 1 for a standard blackbody cell as assumed in the original SQ model, and can be significantly smaller for a PCSC. It should be noted that a large confinement factor (small $f_{esc}$) does not limit the total number of photons escaping the cell per unit time, as the incoming and outgoing photons and electron-hole pairs are assumed to follow the general principle of detailed balance. Importantly, a small $f_{esc}$ reduces the probability (rather than the number) of photons escaping from the surface of the solar cell, so it is possible to establish a high concentration of photons within the cell. Under steady-state conditions, a detailed balance of currents in the solar cells is established (Equation 1), following Kirchhoff's current law and the conservation of particles.

$$J_{\text{sun}} - J_{\text{recom}} - J_{\text{ext}} = 0 \tag{1}$$

where $J_{\text{sun}}$ is the photocurrent density generated by AM1.5G solar irradiation (Equation 2), $J_{\text{recom}}$ is the total recombination current including both radiative and non-radiative recombination currents (Equation 3), and $J_{\text{ext}}$ is the external (output) current density of the solar cell.

$$J_{\text{sun}} = q \int_{E_g}^{\infty} \frac{\lambda}{hc} I_{\text{AM1.5G}} dE \tag{2}$$

where $q$ is the elementary charge, $\lambda$ is the wavelength of the photons, $h$ is the Planks constant, $E$ is the photon energy, $I_{\text{AM1.5G}}$ is the standard AM1.5G spectral irradiance (1 sun), $E_g$ is the bandgap of the photovoltaic absorber material.

$$J_{\text{recom}} = \frac{q f_{\text{esc}}}{\eta_{\text{rad}}} \int_{E_g}^{\infty} \frac{4\pi E^2}{h^3 c^2} \frac{1}{\left[e^{\frac{E-qV}{kT}} - 1\right]} dE \tag{3}$$

where $V$ is the applied voltage corresponding to the quasi-Fermi level splitting in the cell, $f_{\text{esc}}$ is the probability of escape ($0 < f_{\text{esc}} \leqslant 1$) for the luminescent photons, $\eta_{\text{rad}}$ is the efficiency of radiative recombination, $k$ is the Boltzmann constant, and $T$ is the temperature of the solar cell. A derivation of the total recombination current is available in the Supplementary Materials. A full expression of the current-voltage characteristics of the solar cells can be obtained by substituting Equations 2, 3 into Equation 1, yielding

$$J_{\text{ext}} = \int_{E_g}^{\infty} \frac{q\lambda}{hc} I_{\text{AM1.5G}} dE - \frac{q f_{\text{esc}}}{\eta_{\text{rad}}} \int_{E_g}^{\infty} \frac{4\pi E^2}{h^3 c^2} \frac{1}{\left[e^{\frac{E-qV}{kT}} - 1\right]} dE \tag{4}$$

Using Equation 4, the open-circuit voltage ($V_{\text{OC}}$) and short-circuit current density ($J_{\text{SC}}$) of the solar cell can be obtained by setting $J_{\text{ext}} = 0$ and $V = 0$, respectively. The power-conversion efficiencies (PCEs) and fill factors (FFs) of the solar cells can be calculated numerically using Equation 4 following the standard definitions of these metrics (Supplementary Materials).

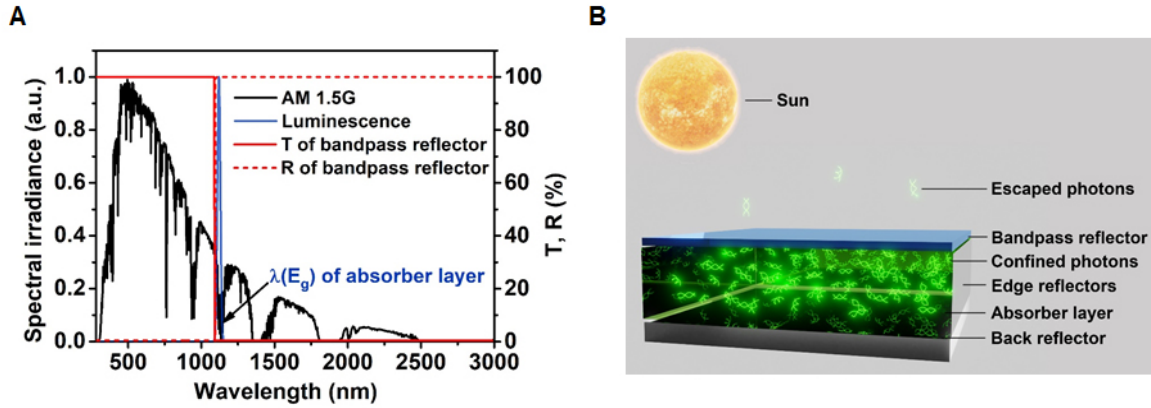

**Figure 1. Optical characteristics of an exemplary photonically-confined solar cell. (A)** The transmission (T, red solid line) and reflection (R, red dashed line) properties of the bandpass reflector for a PCSC, and the luminescence from a model absorber material. Spectral irradiance of AM1.5G solar illumination is also included (black curve). **(B)** An exemplary device structure of a PCSC. The above-bandgap photons from the sun enter the cell via the bandpass reflector and are absorbed by the solar cell absorber layer. The luminescent photons from the absorber are effectively confined within the solar cell by the top bandpass reflector, the back reflector and the edge reflectors, building up a high concentration of photons within the device and a large quasi fermi-level splitting. Under steady-state conditions, a small fraction of internally emitted photons can escape from the top surface, maintaining a detailed balance with the incident photons from the sun and the output current. It should be noted that improved solar cell designs are possible for achieving similar effects.

There might be a question of whether it is possible to have a solar cell for which the probability of absorption does not equal to the probability of escape for a given wavelength. First, the absorption-emission 'reciprocity' does not strictly hold for some systems. For example, there is commonly a Stokes shift between the absorption and luminescence spectra of semiconductors[33,34], suggesting such reciprocity can be broken in these cases. Secondly, it is possible to take advantages of optical designs external to the solar cell functional materials, to allow additional asymmetry in absorption and emission (*vide infra*). However, as we stated above, the general principle of detailed balance, which ensures the reciprocity in the integrated sums of absorbed photons versus outgoing photons and electron-hole pairs, remains to be true in our model. Under open-circuit voltage conditions, the total number of incoming photons equals exactly the total number of emitted photons per unit time, despite there is a considerable built-up of photon concentration within the solar cell.

To illustrate the concept of PCSC in a possible device design, we imagine a solar cell which allows the sunlight above the semiconductor bandgap to be absorbed, while confines all the luminescence (whether it is photo- or electro-luminescence) within the solar cell by, for example, a bandpass reflector with optical characteristics shown in Fig. 1A. It is assumed that, for an ideal semiconductor, the Stokes shift between the near-bandgap absorption and the luminescence, and the bandwidth of the luminescence are sufficiently small. A schematic illustration of how luminescence is restricted in the solar cell is shown in Fig. 1B. An ideal cell is enclosed by top bandpass and back/edge reflectors, so that maximum photon confinement can be achieved. It should be noted that beyond this device structure, there may be alternative designs for the realization of PSCSs.

To model the possible effect of a PCSC, we take halide perovskite, a solar cell materials family that has recently received tremendous attention as an exemplary absorber[20-25,35,36]. Here, the first-generation archetype of the halide perovskite photovoltaic materials, MAPbI$_3$ with a bandgap of 1.55 eV, is used. Halide perovskites are an important example as these simply-prepared materials have been recently shown to produce external quantum efficiencies (or external radiative efficiencies) of over 20% in light-emitting diode applications[37-41].

Our results have shown that for any given radiative efficiency ($\eta_{rad}$), the open-circuit voltage rises substantially as the probability of escape ($f_{esc}$) reduces (Fig. 2A). The maximum theoretical $qV_{OC}$ approaches ~1.55 eV (the bandgap) at a $f_{esc}$ of around $10^{-6}$ for an ideal perovskite with no non-radiative losses ($\eta_{rad} = 1$). A $V_{OC}$ gain of around 0.1V above the maximum $V_{OC}$ predicted by the SQ model may be achieved with a $f_{esc}$ of $10^{-2}$ (1 out of 100 luminescent photons can escape from the cell).

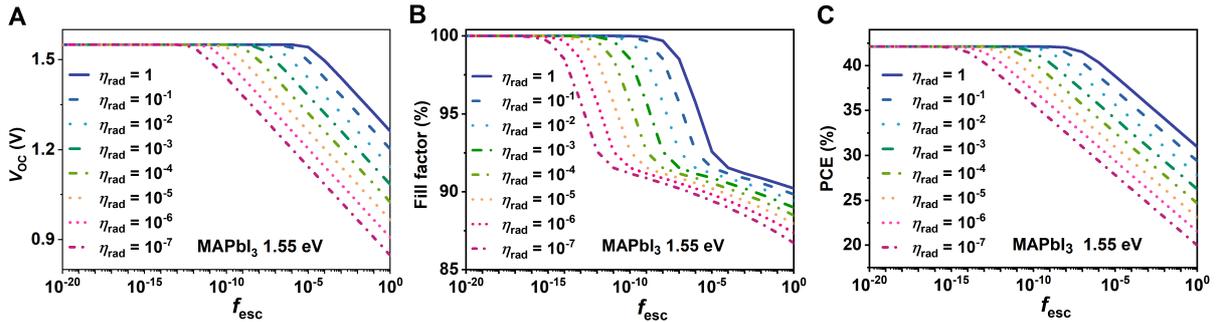

**Figure 2. Modelled photovoltaic performance of a MAPbI$_3$ perovskite-based PCSC, showing the impact of $f_{esc}$ on $V_{OC}$, FF and PCE. (A)** $V_{OC}$ as a function of $f_{esc}$ for a range of radiative efficiencies ($\eta_{rad}$). **(B)** FF as a function of $f_{esc}$ for a range of $\eta_{rad}$. **(C)** PCE as a function of $f_{esc}$ for a range of $\eta_{rad}$.

Another interesting outcome is that the fill factor (FF) of the modelled solar cells approaches 100% as we enhance the photon confinement (lower $f_{esc}$) (Fig. 2B). An example of a modelled IV curve in this scenario is shown in Fig. S1. The very high "squareness" of the diode characteristics may find its application in optoelectronic circuits that demand a very high on-off ratio.

The combined effect of high $V_{OC}$ and FF is the significantly improved PCEs of the solar cells under the photon confinement model. The theoretical maximum PCE of MAPbI$_3$ perovskite solar cells reaches 42.1% (Fig. 2C). It should be noted that all calculations we carried out here are based on a single-junction cell under standard 1-sun illumination, rather than a multiple-junction tandem cell under concentrated sunlight. Another important implication of these results is that even for materials with a very low radiative efficiencies, the maximum theoretical PCE remains to be the same, except that it requires a much lower photon escape factor ($f_{esc}$) to achieve similar efficiencies. In contrast to the original SQ model, non-radiative recombination in a PCSC does not impose restrictions on the upper limits of PCE and $V_{OC}$.

Next, we continue our analysis by examining some of the most representative high-efficiency photovoltaic materials, including silicon (Si) and gallium arsenide (GaAs) (Fig. S2). Similar results can be obtained for these solar cells. The maximum $V_{OC}$ of these solar cells approach their respective bandgaps. The maximum theoretical PCEs are 49% and 45.2% for Si- and GaAs-based PCSCs, respectively.

To provide an overview on how photovoltaic performance changes with varying bandgaps, we computed the theoretical $V_{OC}$, FF limits under the PCSC model ($f_{esc} \ll 1$) and the SQ model ($f_{esc} = 1$, for a blackbody radiator). It can be seen for materials with any given bandgaps that maximum $qV_{OC}$ (where $q$ is elementary charge) approaches the bandgap as we decrease the probability of photon escape (Fig. 3A). This shows that as we suppress the radiative recombination losses that causes a voltage deficit of $E_g$-$qV_{OC}$, the strength of quasi-fermi level splitting allows a substantially higher voltage output compared to what was considered in the SQ model. We note that the state-of-the-art Si, GaAs and perovskite cells produce $V_{OC}$ very close to the $V_{OC}$ limits predicted by the SQ model. Further enhancement of $V_{OC}$ may be possible by taking photon confinement into consideration.

Correspondingly, the theoretical limit of FF approaches 100% for any given bandgaps (Fig. 3B) when $f_{esc}$ for luminescent photons is extremely small. The limit for FF under the SQ model is also computed to provide a comparison. It is interesting to note that the FF of perovskite solar cells is significantly smaller compared to what has been predicted by the SQ model, indicating a relatively large room for improvements in reducing the parasitic losses. Si remains to be the most well-developed solar cell technology considering how close the highest FFs are to the ideal values.

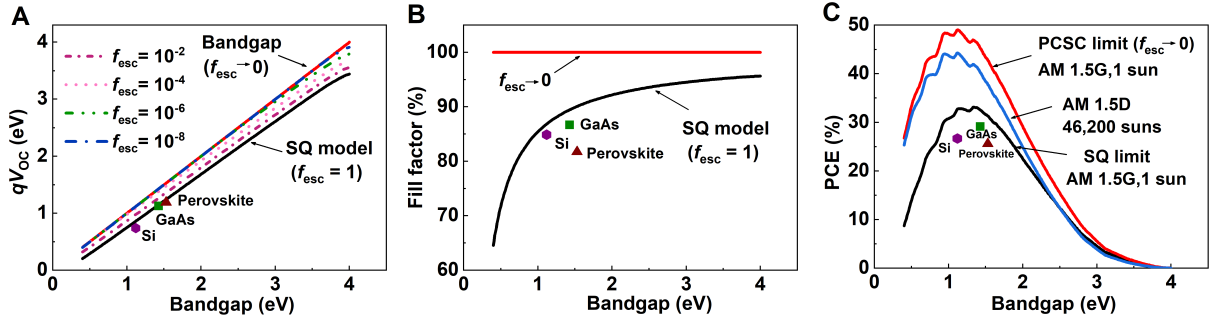

**Figure 3. A summary of photovoltaic performance limits under SQ and PCSC models.** (A) $qV_{OC}$ versus semiconductor bandgap under different $f_{esc}$. The experimental $qV_{OC}$ (data points) of some of the state-of-the-art single-junction solar cells based on Si[42], GaAs[43] and perovskite[25] are shown. (B) Fill factor versus bandgap when $f_{esc} = 1$ (for SQ limit), and $f_{esc} \to 0$ (for a PCSC with an extremely high degree of photon confinement). The experimental FFs (data points) of some of the state-of-the-art single-junction solar cells based on Si[42], GaAs[43] and perovskite[25] are shown. (C) Theoretical PCE limits as a function of bandgap for different models. Results include the SQ limits under 1 sun (AM1.5G) (black curve), under a maximum concentration of 46,200 suns (46200 × AM1.5D) (blue curve), and our PCSC model under AM1.5G (1 sun), $f_{esc} \to 0$ (red curve). The experimental PCEs (data points) of some of the state-of-the-art single-junction solar cells based on Si[42], GaAs[43] and perovskite[25] are shown.

Finally, we compare the theoretical efficiency limits for a range of bandgaps under different theoretical models, including the standard SQ model (1 sun), the SQ model for concentrated solar cells under a maximum concentration of 46,200 suns, and the photon confinement model. We find that the PCSCs have the potential to reach an ultimate efficiency of 49% under 1 sun, substantially higher than the maximum SQ limit of 33.1%, and even higher than the theoretical efficiency limit of 44.3% for single-junction solar cells under the highest possible solar concentration. It is interesting to note that the global maximum efficiency for the photon confinement model appears at a bandgap of 1.12 eV (the $E_g$ of Si), highlighting the strong long-term potential for ultrahigh-efficiency silicon photovoltaic technologies if the photon confinement scheme can be fully explored in the future. It is noted that there is a scope for improvement for Si, GaAs, perovskite and other PV technologies to move further to exceed the conventional theoretical constraints.

In summary, we have presented in our theoretical models that the fundamental efficiency limits for single-junction solar cells under the assumption of photon confinement may potentially exceed the SQ limits. The 1-sun efficiency limits for Si, GaAs and MAPbI$_3$ perovskite-based PCSCs are 49%, 45.2% and 42.1% respectively, significantly higher than the SQ limits (32.7%, 32.5% and 31%) of these materials. The enhancement arises from the suppression of radiative recombination losses, a channel that considered completely inevitable in the classical SQ theory. Under the photon confinement model, $V_{OC}$ values are expected to approach the semiconductor bandgaps, presenting a route to overcoming the long-standing challenge of "voltage deficit" in solar cells. The FFs may approach unity to allow ideal diode characteristics with very high "squareness", a desirable property for optoelectronic operation that requires high on-off ratios. The PCSC concept offers a new pathway to achieving ultra-high efficiencies with the relative simplicity of single-junction cells. Some of the recently reported record-high 1-sun $V_{OC}$ values and PCEs from single-junction cells[3,9,22-25,29,42-45], and experimental results that might exceed the SQ limits in the future, may be partly explained by our model. This is because solar cells with a large dielectric constants and surface texture tend to trap luminescent photons inside the cell, resulting in a small $f_{esc}$, so the possible effects of photon confinement should be revisited in efficiency calculations. While reaching ideal efficiencies in practice requires significant improvements in the optical design of devices, we anticipate our new model to broaden the theoretical horizon for improving solar cell performance beyond the conventional limits.

**Methods**

**Modelling of photovoltaic performance**

The current-voltage characteristics, $V_{OC}$, $J_{SC}$, PCEs and FFs are calculated numerically following the calculations and analyses detailed in the Supplementary Materials, using a MATLAB software R2017b. Standard one-sun irradiation (AM1.5G spectrum) and a solar cell temperature of 298.15K are assumed for all calculations. For standard Si, GaAs and MAPbI$_3$ perovskite solar cells, semiconductor bandgaps of 1.12, 1.43 and 1.55 eV are assumed. The model is applicable to arbitrary solar cell geometries and does not consider Purcell effects. For standard SQ model calculations, ideal "blackbody-like" solar cells with no back reflectors (equivalent to the case of $f_{esc} = 1$ and $\eta_{rad} = 1$) are assumed. Our results of the SQ limits for different bandgaps are in excellent agreement with literature values[2].

**References**

1. W. Shockley, H. J. Queisser, Detailed Balance Limit of Efficiency of p-n Junction Solar Cells. *J. Appl. Phys.* **32**, 510-519 (1961).
2. S. Rühle, Tabulated values of the Shockley–Queisser limit for single junction solar cells. *Sol Energy* **130**, 139-147 (2016).
3. M. A. Green, E. Dunlop, J. Hohl‐Binger, M. Yoshita, X. Hao, Solar cell efficiency tables (version 57). *Prog. Photovolt. Res. Appl.* **29**, 3-15 (2021).
4. D. M. Powell et al., Crystalline silicon photovoltaics: a cost analysis framework for determining technology pathways to reach baseload electricity costs. *Energy Environ. Sci.* **5**, 5874-5883 (2012).
5. F. Dimroth, T. N. D. Tibbits, M. Niemeyer, F. Predan, T. Signamarcheix, Four-Junction Wafer-Bonded Concentrator Solar Cells. *IEEE J. Photovolt.* **6**, 1-7 (2015).
6. S. Essig et al., Raising the one-sun conversion efficiency of III–V/Si solar cells to 32.8% for two junctions and 35.9% for three junctions. *Nat. Energy* **2**, 17144 (2017).
7. A. Polman, M. Knight, E. C. Garnett, B. Ehrler, W. C. Sinke, Photovoltaic materials: Present efficiencies and future challenges. *Science* **352**, aad4424 (2016).
8. M. A. Green, S. P. Bremner, Energy conversion approaches and materials for high-efficiency photovoltaics. *Nat. Mater.* **16**, 23-34 (2017).
9. P. K. Nayak, S. Mahesh, H. J. Snaith, D. Cahen, Photovoltaic solar cell technologies: analysing the state of the art. *Nat. Rev. Mater.* **4**, 269-285 (2019).
10. W. K. Metzger et al., Exceeding 20% efficiency with in situ group V doping in polycrystalline CdTe solar cells. *Nat. Energy* **4**, 837-845 (2019).
11. V. Bermudez, A. Perez-Rodriguez, Understanding the cell-to-module efficiency gap in Cu(In,Ga)(S,Se)$_2$ photovoltaics scale-up. *Nat. Energy* **3**, 466-475 (2018).
12. C. Yan et al., Cu$_2$ZnSnS$_4$ solar cells with over 10% power conversion efficiency enabled by heterojunction heat treatment. *Nat. Energy* **3**, 764-772 (2018).
13. J. Tang et al., Colloidal-quantum-dot photovoltaics using atomic-ligand passivation. *Nat. Mater.* **10**, 765-771 (2011).
14. L. Meng et al., Organic and solution-processed tandem solar cells with 17.3% efficiency. *Science* **361**, 1094-1098 (2018).
15. Y. Cui et al., Over 16% efficiency organic photovoltaic cells enabled by a chlorinated acceptor with increased open-circuit voltages. *Nat. Commun.* **10**, 2515 (2019).

16. D. N. Congreve et al., External Quantum Efficiency Above 100% in a Singlet-Exciton-Fission–Based Organic Photovoltaic Cell. *Science* **340**, 334-337 (2013).

17. J. Hou, O. Inganäs, R. H. Friend, F. Gao, Organic solar cells based on non-fullerene acceptors. *Nat. Mater.* **17**, 119-128 (2018).

18. B. O'Regan, M. Grätzel, A low-cost, high-efficiency solar cell based on dye-sensitized colloidal $TiO_2$ films. *Nature* **353**, 737-740 (1991).

19. M. Freitag et al., Dye-sensitized solar cells for efficient power generation under ambient lighting. *Nat. Photonics* **11**, 372-378 (2017).

20. H.-S. Kim et al., Lead iodide perovskite sensitized all-solid-state submicron thin film mesoscopic solar cell with efficiency exceeding 9%. *Sci. Rep.* **2**, 591 (2012).

21. M. A. Green, A. Ho-Baillie, H. J. Snaith, The emergence of perovskite solar cells. *Nat. Photonics* **8**, 506-514 (2014).

22. Q. Jiang et al., Surface passivation of perovskite film for efficient solar cells. *Nat. Photonics* **13**, 460-466 (2019).

23. M. Jeong et al., Stable perovskite solar cells with efficiency exceeding 24.8% and 0.3 -V voltage loss. *Science* **369**, 1615-1620 (2020).

24. J. J. Yoo et al., Efficient perovskite solar cells via improved carrier management. *Nature* **590**, 587-593 (2021).

25. J. Jeong et al., Pseudo-halide anion engineering for α-FAPbI3 perovskite solar cells. *Nature* **592**, 381-385 (2021).

26. O. D. Miller, E. Yablonovitch, S. R. Kurtz, Strong internal and external luminescence as solar cells approach the Shockley–Queisser limit. *IEEE J. Photovolt.* **2**, 303-311 (2012).

27. L. M. Pazos-Outón et al., Photon recycling in lead iodide perovskite solar cells. *Science* **351**, 1430-1433 (2016).

28. W. Nie et al., High-efficiency solution-processed perovskite solar cells with millimeter-scale grains. *Science* **347**, 522-525 (2015).

29. D. Luo et al., Enhanced photovoltage for inverted planar heterojunction perovskite solar cells. *Science* **360**, 1442-1446 (2018).

30. A. Polman, H. A. Atwater, Photonic design principles for ultrahigh-efficiency photovoltaics. *Nat. Mater.* **11**, 174-177 (2012).

31. R. Lin et al., Monolithic all-perovskite tandem solar cells with 24.8% efficiency exploiting comproportionation to suppress Sn(ii) oxidation in precursor ink. *Nat. Energy* **4**, 864-873 (2019).

32. J. Xu et al., Triple-halide wide–band gap perovskites with suppressed phase segregation for efficient tandems. *Science* **367**, 1097-1104 (2020).

33. F. Yang, M. Wilkinson, E. J. Austin, K. P. O'Donnell, Origin of the Stokes shift: A geometrical model of exciton spectra in 2D semiconductors. *Phys. Rev. Lett.* **70**, 323-326 (1993).

34. Y. Guo et al., Dynamic emission Stokes shift and liquid-like dielectric solvation of band edge carriers in lead-halide perovskites. *Nat. Commun.* **10**, 1175 (2019).

35. A. Kojima, K. Teshima, Y. Shirai, T. Miyasaka, Organometal halide perovskites as visible-light sensitizers for photovoltaic cells. *J. Am. Chem. Soc*. **131**, 6050-6051 (2009).

36. M. M. Lee, J. Teuscher, T. Miyasaka, T. N. Murakami, H. J. Snaith, Efficient hybrid solar cells based on meso-superstructured organometal halide perovskites. *Science* **338**, 643-647 (2012).

37. B. Zhao et al., High-efficiency perovskite–polymer bulk heterostructure light-emitting diodes. *Nat. Photonics* **12**, 783-789 (2018).

38. T. Chiba et al., Anion-exchange red perovskite quantum dots with ammonium iodine salts for highly efficient light-emitting devices. *Nat. Photonics* **12**, 681-687 (2018).

39. K. Lin et al., Perovskite light-emitting diodes with external quantum efficiency exceeding 20 per cent. *Nature* **562**, 245-248 (2018).

40. Y. Cao et al., Perovskite light-emitting diodes based on spontaneously formed submicrometre-scale structures. *Nature* **562**, 249-253 (2018).

41. W. Xu et al., Rational molecular passivation for high-performance perovskite light-emitting diodes. *Nat. Photonics* **13**, 418-424 (2019).

42. K. Yoshikawa et al., Silicon heterojunction solar cell with interdigitated back contacts for a photoconversion efficiency over 26%. *Nat. Energy* **2**, 17032 (2017).

43. A. Richter *et al.*, n-Type Si solar cells with passivating electron contact: Identifying sources for efficiency limitations by wafer thickness and resistivity variation. *Sol. Energy Mater Sol. Cells* **173**, 96-105 (2017).

44. B. M. Kayes et al., 27.6% conversion efficiency, a new record for single-junction solar cells under 1 sun illumination. In Proceedings of the 37th IEEE Photovoltaic Specialists Conference (PVSC), 4-8 (2011).

45. D. Luo, R. Su, W. Zhang, Q. Gong, R. Zhu, Minimizing non-radiative recombination losses in perovskite solar cells. *Nat. Rev. Mater.* **5**, 44-60 (2020).

**Acknowledgements**

This work was supported by the National Key Research and Development Program of China (grant no. 2018YFB2200401), the National Natural Science Foundation of China (NSFC) (61975180, 62005243), Kun-Peng Programme of Zhejiang Province (D.D.), the Natural Science Foundation of Zhejiang Province (LR21F050003), the Fundamental Research Funds for the Central Universities (2019QNA5005, 2020QNA5002), Zhejiang University Education Foundation Global Partnership Fund, and the Engineering and Physical Science Research Council (EPSRC). A.K. acknowledges the Cambridge Nehru Trust, the Cambridge Bombay Society Fund, Trinity-Henry Barlow Scholarship, Haidar Scholarship, and Rana Denim Pvt. Ltd. for financial support. The authors thank Prof. Neil Greenham for useful discussions.

**Author contributions**

D.D. conceived the theoretical framework and planned the project. Q.Z. and A.K. modelled the solar cell performance under the guidance of D.D. D.D. and Q.Z. wrote the paper. Y.L., B.Z. and R.H.F. provided useful discussions and commented on the paper.

**Competing interests**

The authors declare no competing interests.

**Corresponding author**

Dawei Di (daweidi@zju.edu.cn)

Supplementary Materials for

**Photonically-confined solar cells: prospects for exceeding the Shockley-Queisser limit**

Qian Zhou[1], Arfa Karani[2], Yaxiao Lian[1], Baodan Zhao[1,2], Richard H. Friend[2], Dawei Di[1,2]*

3. State Key Laboratory of Modern Optical Instrumentation, College of Optical Science and Engineering; International Research Center for Advanced Photonics, Zhejiang University, Hangzhou, 310027, China.
4. Cavendish Laboratory, University of Cambridge, JJ Thomson Avenue, Cambridge, CB3 0HE, United Kingdom.

* Corresponding author, Email: daweidi@zju.edu.cn

**Supplementary Figures**

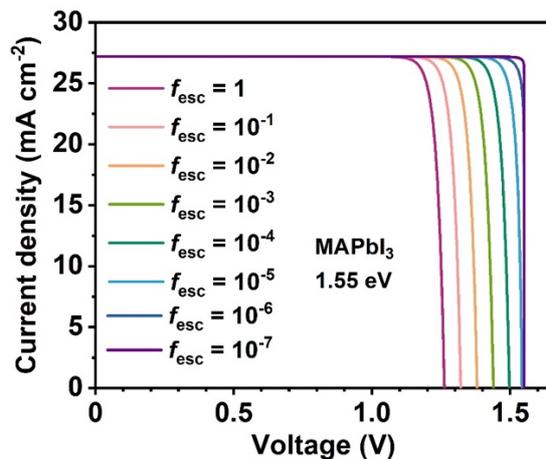

**Figure S1** Current-voltage characteristics for ideal MAPbI$_3$ perovskite solar cells with different $f_{esc}$. The radiative efficiency ($\eta_{rad}$) is assumed to be 1 (100%) for these calculations. The FFs of the solar cells approach unity as $f_{esc}$ becomes very small.

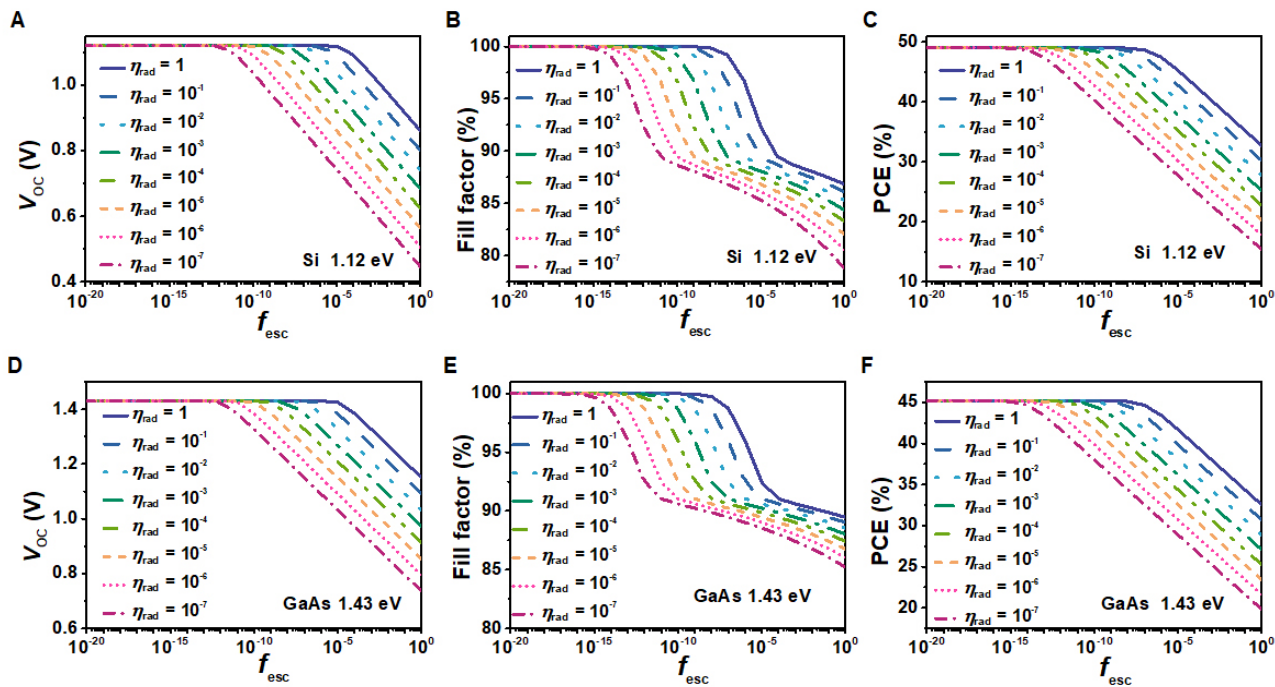

**Figure S2** Modelled photovoltaic performance of Si and GaAs-based PCSCs, showing the impact of $f_{esc}$ on $V_{OC}$, FF and PCE. (A) $V_{OC}$ as a function of $f_{esc}$ for a range of radiative efficiencies ($\eta_{rad}$) for a Si PCSC. (B) FF as a function of $f_{esc}$ for a range of $\eta_{rad}$ for a Si PCSC. (C) PCE as a function of $f_{esc}$ for a range of $\eta_{rad}$ for a Si PCSC. (D) $V_{OC}$ as a function of $f_{esc}$ for a range of radiative efficiencies ($\eta_{rad}$) for a GaAs PCSC. (E) FF as a function of $f_{esc}$ for a range of $\eta_{rad}$ for a GaAs PCSC. (F) PCE as a function of $f_{esc}$ for a range of $\eta_{rad}$ for a GaAs PCSC.

**Supplementary Text**

**Further details of the PCSC model:**

The photocurrent density $J_{\text{sun}}$ generated by AM1.5G solar irradiation can be described by:

$$J_{\text{sun}} = q \int_{E_g}^{\infty} \frac{\lambda}{hc} I_{\text{AM1.5G}} dE \tag{S1}$$

where $q$ is the elementary charge, $\lambda$ is the wavelength of the photons, $h$ is the Planks constant, $E$ is the photon energy, $I_{\text{AM1.5G}}$ is the standard AM1.5G spectral irradiance (1 sun). $E_g$ is the bandgap of the photovoltaic absorber material. The incident power density from the sun ($P_{\text{sun}}$) is ~1000 W/m$^2$, obtained by integrating the spectral irradiance of AM1.5G.

By using generalized Planck law and Kirchhoff's law, together with the consideration of photon confinement, radiative recombination photon flux $\phi_{\text{rad}}$ from the solar cell as a function of external applied voltage can be written as:

$$\phi_{\text{rad}} = f_{esc} \int_{E_g}^{\infty} \frac{2\pi E^2}{h^3 c^2} \frac{1}{\left[e^{\frac{E-qV}{kT}} - 1\right]} dE \tag{S2}$$

where $E$ is the photon energy, $V$ is the applied voltage corresponding to the quasi-Fermi level splitting in the cell, $f_{esc}$ is the probability of escape ($0 < f_{esc} \leq 1$) for the emitted photons, $k$ is the Boltzmann constant, and $T$ is the temperature of solar cell (For the exemplary solar cell design shown in Fig. 1A, $f_{esc} = 1$ for photon energies greater than $E_g + \Delta E$, and $0 < f_{esc} \leq 1$ for regions equal to or smaller than $E_g + \Delta E$, where $\Delta E$ represents the effective linewidth of the luminescence spectrum. This may also slightly shift the absorption onset in Equation S1 in a similar fashion, resulting in a small loss of photocurrent. In the limiting case where $\Delta E$ is sufficiently small for ideal semiconductors, $E_g + \Delta E$ approaches $E_g$ and the loss of photocurrent due to the bandpass filter approaches zero. It should be noted that such limiting conditions are not necessarily required if alternative solar cell designs are implemented). Therefore, the radiative recombination current density ($J_{\text{rad}}$) can be found (Equation S3).

$$J_{\text{rad}} = q f_{esc} \int_{E_g}^{\infty} \frac{2\pi E^2}{h^3 c^2} \frac{1}{\left[e^{\frac{E-qV}{kT}} - 1\right]} dE \tag{S3}$$

The expression for recombination current can be generalized further to include non-radiative recombination. If the efficiency of radiative recombination is assumed to be $\eta_{\text{rad}}$, the relationship between the radiative recombination current $J_{\text{rad}}$ and the total recombination current $J_{\text{recom}}$ including radiative and non-radiative components can be established (Equation S4).

$$J_{\text{rad}} = \eta_{\text{rad}} J_{\text{recom}} \tag{S4}$$

So the total recombination current density $J_{\text{recom}}$ can be written as

$$J_{\text{recom}} = \frac{1}{\eta_{\text{rad}}} J_r = \frac{qf_{\text{esc}}}{\eta_{\text{rad}}} \int_{E_g}^{\infty} \frac{2\pi E^2}{h^3 c^2} \frac{1}{\left[e^{\frac{E-qV}{kT}} - 1\right]} dE \tag{S5}$$

Under steady-state conditions, a detailed balance of currents in the solar cells is established (Equation S6), following Kirchhoff's current law and the conservation of particles.

$$J_{\text{sun}} - J_{\text{recom}} - J_{\text{ext}} = 0 \tag{S6}$$

This leads to a full expression of external current $J_{ext}$ by using Equations S1 and S5.

$$J_{\text{ext}} = q \int_{E_g}^{\infty} \frac{\lambda}{hc} I_{\text{AM1.5G}} dE - \frac{qf_{\text{esc}}}{\eta_{\text{rad}}} \int_{E_g}^{\infty} \frac{2\pi E^2}{h^3 c^2} \frac{1}{\left[e^{\frac{E-qV}{kT}} - 1\right]} dE \tag{S7}$$

The open-circuit voltage ($V_{OC}$) can be calculated by setting $J_{ext} = 0$ in Equation S7, yielding

$$q \int_{E_g}^{\infty} \frac{\lambda}{hc} I_{\text{AM1.5G}} dE - \frac{qf_{\text{esc}}}{\eta_{\text{rad}}} \int_{E_g}^{\infty} \frac{2\pi E^2}{h^3 c^2} \frac{1}{\left[e^{\frac{E-qV_{oc}}{kT}} - 1\right]} dE = 0 \tag{S8}$$

The short-circuit current density ($J_{SC}$) can be obtained by setting $V = 0$ in Equation S7.

$$J_{\text{SC}} = q \int_{E_g}^{\infty} \frac{\lambda}{hc} I_{\text{AM1.5G}} dE - \frac{qf_{\text{esc}}}{\eta_{\text{rad}}} \int_{E_g}^{\infty} \frac{2\pi E^2}{h^3 c^2} \frac{1}{\left[e^{\frac{E}{kT}} - 1\right]} dE \tag{S9}$$

For photons from the sun that can be converted by a typical semiconductor, $E \gg kT$. This leads to $e^{\frac{E}{kT}} \gg 1$, making the second term $\frac{qf_{\text{esc}}}{\eta_{\text{rad}}} \int_{E_g}^{\infty} \frac{2\pi E^2}{h^3 c^2} \frac{1}{\left[e^{\frac{E}{kT}} - 1\right]}$ diminishingly small. This means that $J_{\text{SC}}$ is a constant independent of $f_{\text{esc}}$ and $\eta_{\text{rad}}$.

$$J_{\text{SC}} \approx q \int_{E_g}^{\infty} \frac{\lambda}{hc} I_{\text{AM1.5G}} dE \tag{S10}$$

The efficiency of the solar cell can be calculated numerically using Equation S7, based on the standard definition for PCE.

$$\text{PCE} = \frac{P_{\max}}{P_{\text{sun}}} = \frac{V_{\text{mp}} J_{\text{mp}}}{P_{\text{sun}}} \tag{S11}$$

where $V_{\text{mp}}$ and $J_{\text{mp}}$ are the voltage and current density at the maximum power point, respectively.

The calculation of the fill factor (FF) follows the standard definition.

$$\text{FF} = \frac{V_{\text{mp}} J_{\text{mp}}}{V_{\text{oc}} J_{\text{sc}}} \tag{S12}$$